\documentclass[11pt, a4paper]{article}
\usepackage{amssymb}
\usepackage{eurosym}
\usepackage{amsfonts}
\usepackage{amsmath}
\usepackage{array}
\usepackage{calc}
\usepackage{array}
\usepackage{color}
\usepackage{indentfirst}
\usepackage{graphicx}
\usepackage{hyperref}

\title{The first law of black hole thermodynamics for Taub-NUT spacetime}
\date{\today}

\begin{document}
\author{Remigiusz Durka \thanks{remigiusz.durka@uwr.edu.pl},\\
	%EndAName
	{\small \textit{Institute for Theoretical Physics}, University of Wroclaw,}\\
	{\small Pl.\ Maksa Borna 9, Pl--50-204 Wroclaw, Poland}}
\maketitle

\begin{abstract}
	    
This article focuses on the new approach to the nature of a \textit{nut} parameter $n$, usually treated as a dual charge to the mass. Instead of thinking about it as a gravitational analog of the magnetic monopole, we will show that the Taub--NUT spacetime conceals a peculiar rotation coming from the Misner strings depending on the \textit{nut} value. Besides reconciling a number of confusing results present in the literature, this allows us to establish a consistent description of the black hole thermodynamics for the Lorentzian Taub--NUT spacetime with essential contribution from the Misner strings to the angular momentum and total entropy.  

\end{abstract}

\section{Introduction}

Full understanding of the Taub--NUT spacetime, after more than 50 years since its derivation \cite{Taub:1950ez, Newman:1963yy}, is still quite troublesome. It is not only challenging on the computational level but also conceptually due to the presence of some undesired features. The standard (Euclidean) way of dealing with this metric \cite{Misner:1963fr, Clarkson:2005mc, Hawking:1998ct, Mann:1999pc, Mann:1999bt, Mann:2004mi, Astefanesei:2004kn, Liko:2011cq} unavoidably degenerates the system, eventually neglecting all the interesting and relevant details. Moreover, this still leads to many inconsistencies. 

In this article we are going to deal with the interesting interpretation of $n$ as the one essentially accounting for the rotation. This has a potential to cure many issues and complete the first black hole thermodynamics law, explained now in terms of the angular momentum and entropy associated with the rotating Misner strings.

Proper handling of the Taub--NUT metric should start with the separation of the Lorentzian solution into three cases corresponding to different localization of the string/wire singularity stretching from the black hole horizon to the infinity. By the introduction of the parameter $C=\left\{-1,0,1\right\}$, we obtain, respectively, string going down from the south pole, strings on both poles, and the string going up \cite{Portugues:2009oya, Clement:2015cxa}. The metric itself is given by
\begin{equation}
ds^{2}=-f(r)^{2}(dt-2n\left(  \cos\theta+C\right)  \,d\varphi)^{2}%
+\frac{dr^{2}}{f(r)^{2}}+(n^{2}+r^{2})\,d\Omega^{2}\,,\label{NUT-metric}
\end{equation}
where we define
\begin{equation}
f(r)^{2}=\frac{r^{2}-2GMr-n^{2}}{n^{2}+r^{2}}\,,\label{f1}
\end{equation}
or in the case of the (A)dS with the cosmological constant $\Lambda=\mp\frac{3}{\ell^{2}}$:
\begin{equation}
f(r)^{2}=\frac{r^{2}-2GMr-n^{2}-(r^{4}+6n^{2}r^{2}-3n^{4})\frac{\Lambda}{3}
}{\,{n^{2}+r^{2}}}\,.\label{f2}
\end{equation}
In the limit $n\rightarrow0$ we restore the standard Schwarzschild solution. Presented metric has no curvature singularities, i.e. there is no problem with the Kretschmann scalar,
\begin{align}
K=\frac{8 \Lambda^{2}}{3}
+\frac{48}{(r^2+n^2)^6}\left(n\left(n^{2}-3 r^{2}\right)\left(1-\frac{4 n^{2} \Lambda}{3}\right)n+r\left(3 n^{2}-r^{2}\right)m\right)^2\\
-\frac{48}{(r^2+n^2)^6}\left(n\left(n^{2}-3 r^{2}\right)m - r\left(3 n^{2}-r^{2}\right)\left(1-\frac{4 n^{2} \Lambda}{3}\right) n\right)^2\,,\label{Kretschmann}
\end{align}
for either $r \to 0$ or $r\to r_{H}$. On the other hand, for the angle $\theta=0$ or $\theta=\pi$ (with specific values of $C$ parameter), the metric \eqref{NUT-metric} fails to be invertible. Places where the inverse metric blows up are treated as the wire singularities. Usually, they are considered to be some kind of coordinate artifact without assigning any "physical" meaning to them. Shortly after the publication \cite{Newman:1963yy} delivering considered metric, C.~W.~Misner proposed to remove the wire singularities \cite{Misner:1963fr} by using two separate Euclidean metric patches (with the Wick rotated $t$ coordinate and value of $n\to i\,n$), behaving regularly on the opposite poles, and gluing \textit{good} pieces of two metrics (north part of $C=-1$ and south part of $C=1$). It reminded the Dirac monopole procedure, therefore the wire singularities were called the Misner strings \cite{Misner:1963fr, Clarkson:2005mc, Hawking:1998ct, Mann:1999pc}.
\begin{figure}[h]
	\centering
	\includegraphics[width=0.75\linewidth]{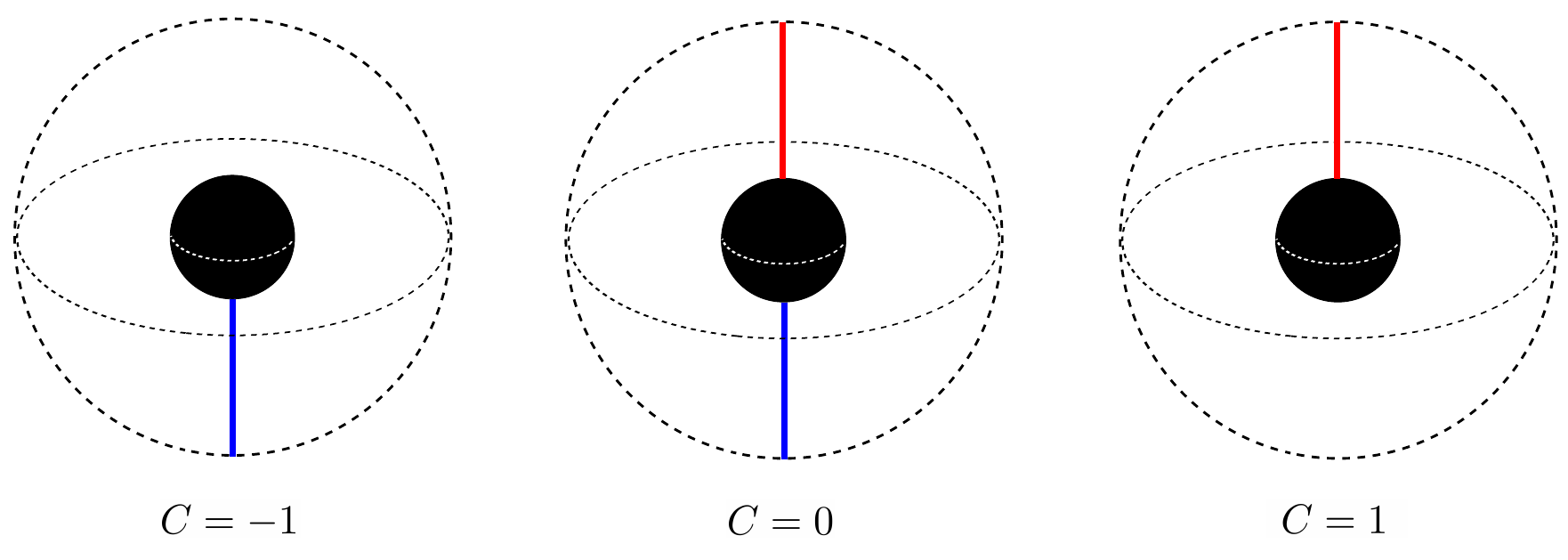}
	\caption[]{Location of the Misner strings}
	\label{fig:taub}
\end{figure}

\noindent\noindent To eliminate them, one has to impose time periodicity condition $\beta=8\pi n$ ensuring the metric regularity. Unfortunately, this quickly becomes a road to nowhere, as we immediately face the number of constraints on the parameters and, in a process, lose all the important features of the new description. Among other things, this translates into fixing the surface gravity to $\kappa=\frac{1}{4n}$ instead of $\kappa=\frac{1}{2}\frac{\partial f(r)^2}{\partial r}|_{r_{+}}=\frac{1}{2r_{+}}$ or the value $\kappa=\frac{1}{2r_{+}}(1-\left(n^{2}+r_{+}^{2}\right)\Lambda)$ \cite{Kouwn:2009qb} and forcing the relationship between the radius $r$ and $n$ value. On the top of that, there is an emergence of some subsystems called the Taub--NUT and Taub--Bolt solutions \cite{Johnson:2014pwa, Fatibene:1999ys, Clarkson:2002uj, Lee:2014tma}, relating differently $n$ with $M$.

\section{From NUT parameter to rotation}

The publication of the metric \cite{Newman:1963yy} (shown to be a coordinate extension of the solution given by Taub \cite{Taub:1950ez}), interestingly not only precedes the discovery of Kerr's solution but is inherently related to it.

Initially, a paper of Ezra T. Newman, Louis A. Tamburino, and Theodore W. Unti, whose initials form the "NUT", had besides the introduction of the stationary, axisymmetric, homogeneous space, a second part corresponding to the discussion of a rotation. Unfortunately, a relation $n=-n$ obtained due to the sign error in the set of equations has led to a claim that the particular class of twisting metrics does not exist. After the error was pointed out in the peer-reviewing process, the second half of the article was discarded \cite{Adamo:2014baa}. Independently from the authors, the reviewer (being none other than Roy Kerr) has pushed the subject of rotation in General Relativity a little bit further and shortly after discovered his famous rotating black hole solution.

Although the eliminating wire singularity employed in 1963 has not resolved many serious issues, Misner's treatment dominated in the literature. It should be recognized that W.~B.~Bonnor, in a rather unknown publication \cite{Bonnor:1969fk} from 1969, was the first person who saw in the semi-infinite Misner string the massless source of the finite (per unit length) angular momentum. His observation was based on the direct comparison between the linearized Kerr solution and Taub-NUT. Later such approach was picked up by \cite{Mazur:1986gb}, and recently reminded by \cite{Griffiths:2009dfa}. We can also find it in V.~S.~Manko and E.~Ruiz article \cite{Manko:2005nm} playing with the exotic "negative" masses. Additionally, it is worth to mention a duality between the hyperbolic Taub--NUT and ultraspinning Kerr solution \cite{Hennigar:2015gan}, along some prior duality shown in \cite{Turakulov:2001jc,Dadhich:2001sz}, as well as some resemblance concerning the general form of metric \eqref{NUT-metric} and rotating cosmic string \cite{Mazur:1986gb, Griffiths:2009dfa}.

Another straightforward piece of evidence linking \textit{nut} parameter with the rotation (and the Kerr parameter $a$) comes from the comparison between the horizons of (AdS)--Kerr-Newman with (AdS)--Taub--NUT:
\begin{align*}
0=(r^{2}+a^{2})(1+\frac{r^{2}}{\ell^{2}})-2Mr+Q^{2} &=  r^{2}-2Mr+a^{2}%
+\frac{r^{4}+a^{2}r^{2}}{\ell^{2}}+Q^{2}\ \ \ (Kerr-Newman)\\
\ \ \   0 & = r^{2}-2Mr-n^{2}+\frac{r^{4}+6n^{2}r^{2}-3n^{4}}{\ell^{2}%
}\ \ \ \ \ (Taub-NUT)
\end{align*}
While we are here, the form of $f(r)^2$ (given by Eq.~\eqref{f2} or in Eq.~\eqref{f1} often written in Schwarzschild's manner as $f(r)^2=1-\frac{2 G M r + 2n^2}{r^2+n^2}$) is in direct analogy to Kerr's \begin{align}
\frac{\Delta_r}{\rho^2}=\frac{r^{2}-2Mr+a^{2}+\frac{r^{4}+a^{2}r^{2}}{\ell^{2}}}{r^2+a^2\cos^2\theta}\,,
\end{align}
which is not a surprise to ones knowing the Plebanski-Demianski family of solutions \cite{Griffiths:2005qp}.

Ultimately, a key element in unraveling the mystery concealing the nature of the \textit{nut} parameter seems to lie within the notion of angular velocity, which for the Taub--NUT solution is given by:
\begin{equation}
\Omega=-\frac{g_{t\varphi}}{g_{\varphi\varphi}}=\frac{-2nf(r)^{2}(\cos
	\theta+C)}{-4n^{2}f(r)^{2}(\cos\theta+C)^{2}+\left(
	n^{2}+r^{2}\right) \sin^{2}\theta }\,.\label{velocity-general}
\end{equation}
No matter, which $C$ parameter configuration we will choose, the evaluation at the horizon makes angular velocity $\Omega_H=0$, simply due to $f(r_{H})^{2}=0$. Moreover, evaluation at infinity in flat regime leads to $\Omega_{\infty}=0$. Remarkably, an example with $\Lambda\neq0$ (let's say AdS $\Lambda<0$) shows that while we still have $\Omega_{H}=0$, at infinity we will observe $\Omega_{\infty}\neq0$. In fact formula \eqref{velocity-general} shows that everywhere outside the horizon we face quite complicated function depending not only from the value of $n$, $M$, $\Lambda$ but also distance $r$ and angle $\theta$.

If the horizon is not rotating, how is that even possible? We have already hinted that the solution of this apparent puzzle can be reconciled by understanding that rotation comes from the string/strings located on the $z$-axis stretching from the horizon to the infinity. Indeed, by going with the $\theta$ angle to zero or $\pi$, the angular velocity \eqref{velocity-general} becomes constant outside the horizon:
\begin{align}
\Omega\Big|_{{}_{C=+1}^{\theta=0}}=\frac{1}{4n}\qquad\mbox{and}\qquad\Omega\Big|_{{}_{C=-1}^{\theta=\pi}}=-\frac{1}{4n}\,.
\end{align}
Interestingly, we will observe the rotation on the opposite side of the $z$-axis, where the strings are not present. This is just a matter of using l'Hospital's rule, e.g. when $C=1$ $$\displaystyle\Omega\Big|_{{}_{C=+1}^{\theta=\pi}}=\lim_{\theta \to \pi}\frac{\alpha(\cos
	\theta+1)}{\beta(\cos\theta+1)^{2}+\gamma\sin^{2}\theta }=\lim_{\theta \to \pi}\frac{\alpha}{2(\beta+(\beta-\gamma)\cos\theta)}=\frac{\alpha}{2\gamma}$$
This means that we have accordingly:
\begin{align}
\Omega\Big|_{{}_{C=+1}^{\theta=\pi}}=-\frac{n f(r)^2}{r^2+n^2}\qquad\mbox{and}\qquad\Omega\Big|_{{}_{C=-1}^{\theta=0}}=\frac{n f(r)^2}{r^2+n^2}\,.\label{opossite-axis}
\end{align}
It is now clear that  Misner's procedure hides a fatal flaw. The two patches were glued to remove the presence of strings. So, with them supposedly out the picture, we have got a counter-frame-dragging in the opposite hemispheres coming... out of nowhere.

Note, that one could use the unified formula to describe the angular velocity of the strings
\begin{align}
\Omega|_{\theta=\{0,\pi\}}=\frac{1}{2n(\cos\theta+C)}\Big|_{\theta=\{0,\pi\}}\label{angular-velocity}
\end{align}
but we must be aware of the subtlety concerning \eqref{opossite-axis}. With the formula \eqref{angular-velocity} we are on the verge of yet another important discovery. For the case $C=0$, therefore with both strings present, we surprisingly see
\begin{align}
\Omega\Big|_{{}_{C=0}^{\theta=0}}=\frac{1}{2n}\qquad\text{and}\qquad\Omega\Big|_{{}_{C=0}^{\theta=\pi}}=-\frac{1}{2n}\,,
\end{align}
so the two strings are rotating in the opposite directions! This is actually the missing piece, explaining some of the confusing results and opening up new directions of research. We can find a confirmation of such bizarre counter-rotating strings firstly in Manko and Ruiz publication \cite{Manko:2005nm}, and later in the notion of twisted black holes \cite{Ong:2016cbo, Gray:2016pbu}.

\section{Angular momentum}

Some of my past work \cite{Durka:2011zf, Durka:2011yv, Durka:2012wd} concerned calculations of the Noether charges involving the topological terms regularization with a negative cosmological constant \cite{Aros:1999id}. Such method removes the divergence and assures the right factors, all in the solution independent manner. Both mass and angular momentum are resulting from the evaluation of charges at infinity, respectively for the time-like $\partial_t$ and rotational $\partial_\varphi$ Killing vector.

Further generalized formula, accounting for a contribution from the Pontryagin and Holst terms \cite{Durka:2011zf, Durka:2012wd}, has been applied to various metrics, including the Taub-NUT case. For the standard Einstein-Cartan action together with the Euler term, $\partial_t$ delivered the value \begin{align}
\mbox{Mass}=Q[\partial_t]|_\infty=M\,,
\end{align}
as expected. In turn, the addition of the Pontryagin/Holst terms picked up the value:
\begin{align}
\mbox{Dual Mass}=n \left(1-\frac{4 n^{2} \Lambda}{3}\right)\,.\label{dual-mass}
\end{align}
One can look at it as an analogy to the electric/magnetic charges
\begin{align}
q_e=\frac{1}{4\pi} \int_{\partial V} * \mathcal{F} \qquad\mbox{and}\qquad q_{m}=\frac{1}{4\pi} \int_{\partial V} \mathcal{F}\,.
\end{align}
That agrees with treating the \textit{nut} parameter as the gravitational analog of the magnetic monopole \cite{Clarkson:2005mc,Portugues:2009oya, Zee:1986sr, Bossard:2008sw}. Mind, however, that the metric \eqref{NUT-metric} did not come from particular configuration. Instead, it was established merely as the deformation of the Schwarzschild solution with the transformation governed by $n$. Common naming of a single $n$ as the dual charge/mass seems to be misused. It is better to reserve that name to the (composite) expression \eqref{dual-mass} pointed above, which obviously reduces to $n$ for $\Lambda=0$. This is reflected in many other aspects (see for instance the electric/magnetic mass duality within the Kretschmann scalar formula \eqref{Kretschmann}). Note, that the often occurring condition $M=n\left(1-\frac{4 n^{2} \Lambda}{3}\right)$ is nothing else than the self-dual relation \cite{Araneda:2018orn}. 

Analyzed in the same manner $J=Q[\partial_\varphi]|_\infty$ (a charge evaluated at infinity for $\partial_\varphi$, which we associate with the angular momentum) turned out to be zero for the Taub-NUT \cite{ Durka:2011yv, Durka:2012wd}. However, this was only studied in $C=0$ case. Repeating the same calculations for the general form of metric~\eqref{NUT-metric} unexpectedly brings
\begin{align}
J=3\, C\, M n\,.
\end{align}
The numerical factor "3", appearing later also in \cite{Hennigar:2015gan}, comes from usually overlooked subtlety, namely the true rotational Killing $\xi_{\varphi}$ vector for the Taub-NUT metric \cite{Portugues:2009oya} is of the form 
\begin{align}
\xi_{\varphi}=\partial_{\varphi}-2n C\partial_{t}\, ,
\end{align}
and not single $\partial_{\varphi}$, like in the Kerr spacetime. Eventually, with this out the way, the angular momentum get the form of $J=Q[\xi_\varphi]|_\infty=C\, M n$, strikingly resembling the formula $J=M a$ for the Kerr metric.

Leaving some speculations aside for future work, like playing with the parameter $n$ as some sort of the projection of the Kerr parameter $n\sim a \cos\theta$ on the $z$-axis, we will end this section with some interesting observations.  The well-known Dirac quantization condition expressing the product of $q_e$ with $q_m$ as the integer in the units of $\hbar$, by the means of analogy now for the product of $M$ and $n$ \cite{Mazur:1986gb}, would cause the quantization of the angular momentum $J$. Also note, that the angular momentum comes rather strangely as the outcome of mass being associated with the non-rotating black hole, whereas the source of rotation is located elsewhere. 

It goes without saying, that for the solution with two strings, $C=0$, integrating over $\theta$ angle from $0$ to $\pi$ exactly cancels the north and south contributions, thus one ends up with the angular momentum being globally zero $\sum J_{i}=Mn-Mn=0$. This, however, does not mean we will have no impact on the first law of black hole thermodynamics. It is easy to see, that taking the sum of two hemispheres with counter-rotating strings leads to non-vanishing contribution
\begin{align}
\sum\Omega_{i}dJ_{i}=\frac{1}{2n}d\left(  Mn\right)  +\left(  -\frac{1}%
{2n}\right)  d\left(  -Mn\right)=\frac{1}{n}d\left(  Mn\right)\neq0\,.\label{bh-thermo-with-zero-c}
\end{align}

\section{Horizons}

The event horizon can be defined through the Killing horizon, where the horizon generator $\xi=\partial_{t}+\Omega_{H}\partial_{\varphi}$ has a vanishing norm \cite{Carlip:1999cy}. This requires the angular velocity evaluated at the horizon $\Omega_{H}$ but this turned out to be zero, which is in contrast to value of $\Omega_{H}=a\left(  1-\frac{a^{2}}{\ell^{2}}\right)/(r_{H}^{2}+a^{2})$ for the AdS-Kerr metric. Remarkably, within the Taub--NUT spacetime we can consider another "Killing horizon generator", which applies to the the location of the Misner strings
\begin{align}
\xi=\partial_{t}+\Omega_{s}\partial_{\varphi}\, ,\label{Killing_string}
\end{align}
were $\Omega_{s}$ is the angular velocity of the string \eqref{angular-velocity}. Depending on the particular case, $\Omega_s$ is equal $\frac{1}{4n}$, $-\frac{1}{4n}$ or $\pm\frac{1}{2n}$. The norm of \eqref{Killing_string} becomes zero as the $\theta$ angle goes to 0 or $\pi$. We can also use it to evaluate the corresponding surface gravity $\kappa_s$ from the definition $\kappa^2_s=\frac{1}{2}(\nabla_\mu \xi_\nu)(\nabla^\mu \xi^\nu)$. Calculations show that for both the flat and the AdS regimes $\kappa$ equals to $\frac{1}{4|n|}$ for a single string in either $C=\pm1$ cases, and $\frac{1}{2|n|}$ for each of two strings when we consider $C=0$.

Confirmation of such picture can be found in Carlip paper \cite{Carlip:1999cy}, although without pointing the exact (rotational) reason behind the given there $\xi=\partial_{t}+\frac{1}{4n}\partial_{\varphi}$ configuration. His setup with just one string up was seems to be only due to the consistency condition. Interestingly, Carlip also assigned the entropy to the string to take the form of $S_s=\frac{8\pi n}{4G} \int^\infty_{r_{+}} dr$. This might be understood in terms of $S_s=A^{string}/4G$, where $A^{string}$ would be an induced volume evaluated at $\theta=0$ and constant $t$, therefore requiring the integration with respect to  $\sqrt{-h}\,d^2x=|4n|dr\,d\varphi$. 

Now, in the $C=0$ case we would have the half of this value but assigned to two strings
\begin{align}
A^{string}_{up}=A^{string}_{down}=\int_0^{2\pi}\int_{r_{+}}^{\infty}2|n|dr\,d\varphi\,.
\end{align}
This would lead to the cumulative impact on the black hole thermodynamics through $T_s d S_s=T^{up}_s dS^{up}_s+T^{down}_s dS^{down}_s$, which basically it comes down to
\begin{align}
T_s d S_s=2\left(\frac{1}{2\pi}\frac{1}{2|n|}\right) d\left(2\pi\, \frac{2|n|}{4G}\int_{r_{+}}^{\infty}dr\right)=\frac{1}{n}d\left( \frac{n}{2G}\int_{r_{+}}^{\infty}dr\right)\,.\label{Carlip-entropy}
\end{align}

\section{First law of the black hole thermodynamics}

The first law of the black hole dynamics connects the change of mass $dM$ with the change of event horizon area $dA$, angular momentum $dJ$, and electric charge $dQ$, along with the values of the surface gravity $\kappa$, angular velocity $\Omega$, and electric potential $\Phi$:
\begin{align}
dM = \frac{\kappa}{8\pi G}\,dA+\Omega dJ+\Phi dQ\,.
\end{align}
Works of Bekenstein and Hawking gave it the form of the thermodynamic law by associating the entropy with the area of the event horizon, $Entropy=\frac{A}{4G}$, and the temperature with the surface gravity, $T=\frac{\kappa}{2\pi}$. Although the \textit{nut} parameter $n$ is commonly described as the \textit{dual mass} it has not been not satisfactorily incorporated into the thermodynamic description (contrary to the usual electric and magnetic charges \cite{Caldarelli:1999xj}).

The standard procedure for overcoming obstacles in the form of Misner strings somehow forces the "extreme" case $M=n$ (or $M=n\left(  \ell^{2}+4n^{2}\right)/\ell^{2}$ for the AdS) along with other constraining conditions that are strongly degenerating the whole system. That makes it useless in detailed considerations because ultimately the \textit{nut} parameter is only effectively affecting things while being explicitly gone from the formulas. Altogether, with all the juggling between the Nut/Bolt subsystems, it seems that we have only obtained \textit{left-hand side} equal to \textit{right-hand side}, rather than captured the actual first law.

Surprisingly, we are about to see a very simple procedure of deriving the first law for the Lorentzian Taub--NUT spacetime. It all goes down to a statement that, if there is something fundamental about the black hole thermodynamics, it has to be an area of the horizon. By taking its variation and multiplying the surface gravity related to the temperature, the further terms will be obtained in a systematic manner. Equipped with the number of insights, presented in previous sections, we will be able to give it quite a consistent interpretation, although some issues will still remain to be resolved.
  
\subsection{Kerr}

Before we go to the Taub--NUT metric, let us first see a derivation of the first law for the Kerr solution. The horizon area $A=\int\left.  \sqrt{g_{\theta\theta}\,g_{\varphi\varphi}}\right\vert _{r_{+}}d\theta d\varphi=4\pi(r_{+}^2+a^2)$ implies the derivative
\begin{align}
dA  &  =4\pi(2 r_{+} d r_{+}+2ada)\\
\frac{dA}{8\pi}  &  =r_{+}d r_{+} + ada\,.
\end{align}
The horizon definition $r_{+}^{2}-2 M r_{+}+a^{2}=0$ provides $\left(
r_{+}-M\right)  d r_{+}=r_{+} dM-ada$, therefore we can write
\begin{align}
\frac{dA}{8\pi}  &  =r_{+}\frac{r_{+}dM-ada}{\left(r_{+}-M\right)  }+ada\\
&  =\frac{r_{+}^{2}}{r_{+}-M}dM+\left(  1-\frac{r_{+}}{r_{+}-M}\right)  ada\\
&  =\frac{r_{+}^{2}}{r_{+}-M}dM-\left(  \frac{Ma}{r_{+}-M}\right)  da\,.
\end{align}
At the same time the surface gravity $\kappa=\frac{1}{2r_{+}}\,\frac{(r_{+}^{2}-a^{2})}{(r_{+}^{2}+a^{2})}$ reads $\kappa=\frac{r_{+}-M}{r_{+}^{2}+a^{2}}$. Therefore, by multiplying the both sides we obtain
\begin{align}
\frac{\kappa}{2\pi}\frac{dA}{4}  & =\frac{r_{+}^{2}}{r_{+}^{2}+a^{2}}dM-\frac{Ma}{r_{+}^{2}+a^{2}}da\nonumber\\
&  =dM-\frac{a^{2}}{r_{+}^{2}+a^{2}}dM-\frac{Ma}{r_{+}^{2}+a^{2}}da\nonumber\\
&  =dM-\frac{a}{r_{+}^{2}+a^{2}}\left(  adM+Mda\right) \nonumber\\
&  =dM-\frac{a}{r_{+}^{2}+a^{2}}d\left(M a\right)\, . \nonumber
\end{align}
After identifying the horizon temperature $T=\frac{\kappa}{2\pi}$, angular velocity  $\Omega=-\frac{g_{t\varphi}}{g_{\varphi\varphi}}\Big|_{r_{+}}%
=\frac{a}{r_{+}^{2}+a^{2}}$, entropy $S=A/4$, as well as the angular momentum $J=M a$, the first law becomes
\begin{equation}
dM=TdS+\Omega dJ\,.
\end{equation}

\subsection{Taub--NUT}

Presented approach quite universally captures the correct black hole thermodynamics. The same procedure could be applied even to the AdS--Kerr--Newman solution \cite{Caldarelli:1999xj}. Let us now in the same way examine the Taub--NUT solution. We will start with the asymptotically flat case assured by \eqref{f1} with explicit $G$ constant for some better guidance. Through a condition $r_H^{2}-2 M G r_{H}-n^{2}=0$ we obtain two radius solutions 
\begin{equation}
r_{\pm}=GM\pm\sqrt{\left(  GM\right)  ^{2}+n^{2}}\,. \label{radiusTaub}%
\end{equation}
We choose $r_H=r_{+}>0$ as a black hole radius to neglect a negative value (but elsewhere we should keep in mind a possibility of wormhole scenario).
The surface gravity has the same form as in the Schwarzschild solution,
\begin{equation}
\kappa=\left.  \frac{1}{2}\frac{\partial f(r)^{2}}{\partial r}\right\vert
_{r=r_{+}}=\frac{1}{2r_{+}}\, ,
\end{equation}
but this time due to \eqref{radiusTaub} it has a different value depending also on $n$. Differentiation of the horizon area, being now
\begin{equation}
A=\int_0^{2\pi}\int_0^\pi\sqrt{g_{\theta\theta}\,g_{\varphi\varphi}}\Big|_{r_{+}}d\theta d\varphi
=4\pi\left(  r_{+}^{2}+n^{2}\right)\,,
\end{equation}
and multiplying result by $\kappa/G$, produces
\begin{equation}
\kappa\,\frac{dA}{8\pi G}=\frac{1}{2 G}dr_{+}+\frac{1}{2 G}\frac{n}{r_{+}}dn\,.
\end{equation}
Because we want eventually to achieve a total derivative of $M$, the second term is a subject of rewriting%
\begin{align}
\frac{\kappa}{2\pi}\frac{dA}{4 G}  &  =\frac{1}{2 G}d\,r_{+}+\frac{1}{2 G n}\frac
{n^{2}}{r_{+}}dn\\
&  =\frac{1}{2 G}d\, r_{+}+\frac{1}{2G n}d\left(  \frac{n^{3}}{r_{+}}\right)  -\frac
{1}{2 G}d\left(  \frac{n^{2}}{r_{+}}\right) \\
&  =\frac{1}{2 G}d\left( r_{+}-\frac{n^{2}}{r_{+}}\right)  +\frac{1}{2 G n}d\left(  \frac{n^{3}}{r_{+}}\right)  \,.
\end{align}
Using the formula $\frac{1}{2 G r_{+}}\left(  r_{+}^{2}-n^{2}\right)=M$ completes the first law for the Lorentzian Taub--NUT, which written in the usual order becomes
\begin{equation}
dM=TdS-\frac{1}{2 n}d\left(  \frac{n^{3}}{G r_{+}}\right)  \,.\label{first-NUt-flat}
\end{equation}
Let's postpone a discussion concerning the interpretation and now proceed to the derivation of similar formula for the AdS case.

\subsection{Taub--NUT-AdS}

The AdS horizon definition \eqref{f2} is now equipped with an additional $\Lambda=-\frac{3}{\ell^2}$ piece. Differentiation of the area is still given by $\frac{dA}{8\pi}=r_{+}d r_{+}+ndn$, but the surface gravity becomes
\begin{equation}
\kappa=\frac{1}{2r_{+}}\left(  1+\frac{3(n^{2}+r_{+}^2)}{\ell^{2}}\right) \,.
\end{equation}
The horizon area combined with the surface gravity brings
\begin{align}
\kappa\,\frac{dA}{8\pi}  & =\frac{1}{2}\left(  1+\frac{3(n^{2}+r_{+}^2)}{\ell^{2}}\right)  \left(
d r_{+}+\frac{n}{r_{+}}dn\right)\,.
\end{align}
Using $r_{+}^2-n^{2}=2 M r_{+}-\frac{\left(  r_{+}^{4}+6n^{2}r_{+}^2-3n^{4}\right)  }
{\ell^{2}}$, and following similar steps as above, reproduces
\begin{align}
\frac{\kappa}{2\pi}\frac{dA}{4}  &  =dM+\frac{1}{2n}d\left(  \frac{n^{3}}%
{r_{+}}\right)  -d\left(  \frac{\left(  r_{+}^{3}+6n^{2}r_{+}-3\frac{n^{4}}{r_{+}}\right)
}{2\ell^{2}}\right)  +\frac{3(n^{2}+r_{+}^2)}{\ell^{2}}\frac{1}{2}\left(
d r_{+}+\frac{n}{r_{+}}dn\right) \\
&  =dM+\frac{1}{2n}d\left(  \frac{n^{3}}{r_{+}}\right)  +\frac{3}{2 n\ell^{2}%
}\left(  \frac{n^2}{r_{+}}\left(  5n^{2}-3r_{+}^2\right)  dn-\frac{n^{3}}{r_{+}^2}%
(n^{2}+r_{+}^2)d r_{+}\right)  \,.
\end{align}
Last expression can be shown to have a form $\frac{1}{2n}\frac{3}{\ell^{2}}\left(  d\left(  \frac{n^{5}}{r_{+}}\right)
-d\left(  n^{3}r_{+}\right)  \right)$, thus
\begin{equation}
dM=\frac{1}{2\pi}\frac{1}{2r_{+}}\left(  1+\frac{3(n^{2}+r_{+}^{2})}{\ell^{2}%
}\right)  d\left(  \frac{4\pi\left(  r_{+}^{2}+n^{2}\right)  }{4}\right)
-\frac{1}{2n}d\left(  \frac{n^{3}}{ r_{+}}\left(  1-\frac{3\left(  r_{+}%
	^{2}-n^{2}\right)  }{\ell^{2}}\right)  \right) \,.\label{first-NUt-AdS}
\end{equation}
There is subtlety concerning perturbation of the cosmological constant and thermodynamic volume \cite{Dolan:2010ha, MacDonald:2014zaa,Johnson:2014xza}, but it will be addressed elsewhere.

\section{First law}

In the previous section, we have arrived to the two formulas: \eqref{first-NUt-flat} and \eqref{first-NUt-AdS}. Now it is time to give them some thermodynamic interpretation. Contrary to the Euclidean treatment (involving a number of identifications like $M=n$ or $r_{+}=n$), in a taken Lorentzian regime we have accepted metric as it is, to see how $n$ might explicitly modify the first law. In particular, this could translate into a contribution similar to change of mass $dn$, string's angular velocity and change of momentum $\Omega_{s} dJ_{s}$, string's temperature and change of entropy $T_{s} dS_{s}$, or some analog of the potential and change of "charge" $\Phi_n\, dn$. Instead, the exact calculations for the metric with $\Lambda=0$ led to the formula \eqref{first-NUt-flat} with quite unusual $-\frac{1}{2 n G}d\left(  n^{3}/ r_{+}\right)$ contribution. Luckily, realizing that
\begin{align}
\frac{n^{3}}{r_{+}}=\frac{n}{r_{+}}n^{2}=\frac{n}{r_{+}%
}\left(  r_{+}^{2}-2GMr_{+}\right)  =nr_{+}-2GMn\,, 
\end{align}
along knowing the definition of horizon's temperature $T=\frac{1}{2\pi}\frac{1}{2r_{+}}$ and area $A=4\pi\left(r_{+}^{2}+n^{2}\right)$, immediately brings new, and final, form with $n$ and its change explicitly present
\begin{equation}
dM=Td\left(  \frac{A}{4G}\right)  +\frac{1}{n}d\left(  Mn\right)  -\frac
{1}{2n}d\left(  \frac{nr_{+}}{G}\right)\,.\label{first-law-Taub-NUT-flat}
\end{equation}
Remarkably, $M n$ quantity has already been interpreted as the angular momentum, along with the angular velocity of the string proportional to the inverse of $n$. It should be noted, that this agrees perfectly with the heuristic argument given in \eqref{bh-thermo-with-zero-c}.

The last contribution appears to capture, at least partially, the change of string's entropy associated with the Killing horizon \eqref{Killing_string} already found by Carlip \cite{Carlip:1999cy}, along with the temperature proportional to the inverse of $n$, just as it was mentioned in Eq.~\eqref{Carlip-entropy}. Carlip's article took into account the string going to infinity, so the integration was performed up to the infinity, but at the same time, the angular momentum $J$ of the semi-infinite string should do the same.  Note that \cite{Bonnor:1969fk} mentions the finite $J$ per unit of the length, while \cite{Manko:2005nm} introduces the angular momentum contribution going to infinity as $r\to \infty$. Interestingly, a minus sign between the last two terms in \eqref{first-law-Taub-NUT-flat} might reconcile the missing (infinite) contribution of the entropy of the string as it might be canceled by the \textit{more proper} angular momentum value. As the Taub-NUT metric is not exactly asymptotically flat or AdS, we must be more cautious within the boundaries while calculating particular charges. Evaluating things at the standard horizon and at infinity does not cover everything, as one must consider boundary surrounding the Misner strings. Mention here method of regularizing the Noether charges by the topological terms comes with its own issues and possibly requires more careful approach to the Taub-NUT spacetime \cite{Frodden:2017qwh}.

For completeness, we conclude with the non-vanishing cosmological constant $\Lambda=-3/\ell^2$ case. Without going much into the details, ultimately \eqref{first-NUt-AdS} is being rewritten to take the form
\begin{equation}
dM=Td\left(  \frac{A}{4G}\right)  +\frac{1}{n}d\left(  Mn\right)  -\frac
{1}{2n}d\left(  \frac{nr_{+}}{G}\left(  1+\frac{r_{+}^{2}}{\ell^{2}}+\frac{3n^{2}}{\ell^{2}}\right)  \right)\,.
\end{equation}

Obtained expressions definitely require further explanation and putting heuristic arguments in more formalized context. In the end, although all the hints concerning the validity of the presented framework, the proposed form of black hole thermodynamics might be far from being complete and require other interpretation.

There is still a number of unresolved issues. One of the most serious ones corresponds to the place, where the strings come out from the horizon. By making a transition from the notion of "wire singularity" to the Killing horizon associated with the Misner strings, we must deal now with some union of two Killing horizons with two temperatures assigned to them, which might require fixing their values.  

Calling the Taub--NUT metric "a counterexample to almost anything" in General Relativity \cite{Misner:1965zz} perfectly reflects how generous this solution was as the subject of various queries and generalizations. However, all the unresolved, unusual, and unclear results constantly show us how poorly we understand the true structure behind Taub-NUT space-time. The reason might lie in the overused assumptions. As we saw, the contribution from Misner strings (which we have been neglecting for decades) not only explains some of the problems but is essential to obtain the first law of the black hole thermodynamics. In recent years there were significant efforts in demystifying and embracing their existence \cite{Clement:2015cxa, Clement:2015aka}. This also includes works concerning the astronomical observations \cite{Chakraborty:2017nfu, Chakraborty:2019rna} and effects on the black hole shadow \cite{Abdujabbarov:2012bn}. The question,  if \textit{nut} represents verifiable physical observable or is just an interesting formal toy model, is still open. However, the unusual configuration makes one rather pessimistic about the subject.

Additionally, more work has to be done to reconcile results obtained here with the Euclidean approach, as well as consequences for the notion of the "magnetic" monopole showing a \textit{rotational} character. An open question is whether it could teach us something new, not only in the gravity context but also in electromagnetism, where we still do not have any experimental evidence of magnetic monopole existence.

My interest in the subject originated from some unfinished outcomes of the past work and gradual efforts to explain various pathologies and inconsistencies hidden under the hood of Taub-NUT spacetime. Formulating the first law \eqref{bh-thermo-with-zero-c} with the essential role of angular velocity/momentum coming from the string was obtained around 2016/2017. Since then, although not published, the outcomes were fairly publicized on various seminars and talks given throughout Chile (PUCV, CECs), Italy (Torino) and Poland. Recently other authors reached, to some extent, somehow similar range of results and features exploiting different method. Notable advancements concern more detailed handling the boundaries, dyon generalization, as well as exploring the possibility of the thermodynamic volume \cite{Kubiznak:2019yiu, Ballon:2019uha, Bordo:2019tyh, Bordo:2019rhu}. Yet, many aspects still ask for more satisfactory resolution and putting in a better context.

\section{Conclusions}

This article tries to provide new insight into the nature of Lorentzian Taub--NUT spacetime. Surprisingly, not treating $n$ as a gravitomagnetic monopole, but seeing it as accounting for the rotation, allows us to complete the first black hole law, and explain it in terms of angular momentum and entropy associated with the Misner strings. The new approach hints a better understanding of the underlying metric structure with some intriguing consequences when it comes to the concept of "magnetic" counterpart of gravity and establishing (to this day rather incomplete) description of the corresponding black hole thermodynamics.  

\section{Acknowledgment}

This work was supported by Grant for Young Researchers 0420/2716/18. The author would like to thank for the kind hospitality of Chilean institutes in Valpara\'iso (PUCV) and Valdivia (CECs), where some key elements of this work were carried out and discussed during a completion of FONDECYT postdoc and 2018 visit.

\end{document}